\documentclass[preprint]{aastex631}

\newcommand{\Dds}{D_{ds}}
\newcommand{\Dd}{D_{d}}
\newcommand{\Ds}{D_{s}}

\usepackage{amsmath}

\begin{document}

\title{Investigating Extreme Scattering Events by Volumetric Ray-tracing}

\correspondingauthor{Kelvin Au}
\email{auk@myumanitoba.ca}

\author[0000-0002-3908-3822]{Kelvin Au}
\affiliation{Department of Physics and Astronomy \\ University of Manitoba \\
30A Sifton Road \\
Winnipeg, Manitoba, Canada, R3T 2N2}


\author{Jason D. Fiege}
\affiliation{Department of Physics and Astronomy \\ University of Manitoba \\
30A Sifton Road \\
Winnipeg, Manitoba, Canada, R3T 2N2}

\author[0000-0003-2953-2054]{Adam Rogers}
\affiliation{Department of Physics and Astronomy \\ University of Manitoba \\
30A Sifton Road \\
Winnipeg, Manitoba, Canada, R3T 2N2}



\begin{abstract}

Extreme scattering events (ESEs) are observed as dramatic ($>50\%$) drops in flux density that occur over an extended period of weeks to months. Discrete plasma lensing structures are theorized to scatter the radio waves produced by distant sources such as pulsars, causing the signature decrease in flux density and characteristic caustic spikes in ESE light curves. While plasma lens models in the extant literature have reproduced key features of ESE light curves, they have all faced the problem of being highly over-dense and over-pressured relative to the surrounding interstellar medium (ISM) by orders of magnitude. We model ESEs by numerically ray-tracing through analytic, volumetric plasma lens models by solving the eikonal equation. Delaunay triangulation connecting the rays approximates the wavefront, generating a mapping from the observer plane to the source plane to account for multiple-imaging. This eikonal method of ray-tracing is tested against known analytic solutions and is then applied to a three-dimensional Gaussian-distributed electron volume density lens, and a filament model inspired by \cite{Grafton2023MNRAS.522.1575G}. We find convergence of our numerical results with established analytic solutions validating our numerical method, and reproduce ESE-like light curves. Our numerical ray-tracing method lends itself well to exploring the lensing effects of volumetric turbulence as well as sheet-like lenses, which is currently in progress.

\end{abstract}



\section{Introduction}
\label{sec:Introduction}
Distant, highly energetic objects such as pulsars and quasars are luminous and emit strongly in the radio regime. Typically, the mechanism by which the radio radiation arises comes from synchrotron radiation. Pulsars and quasars possess extremely strong magnetic fields; $\sim 10^{9} \, \text{G}$ for pulsars, and up to $\sim 10^{6} \, \text{G}$ in quasar jets. These strong magnetic fields accelerate free electrons, which emit radiation in the radio regime. Radiation is influenced by free electrons in plasma as it propagates through space. The influence on the propagation of radiation at frequency $f$ through cold plasma is characterized by an index of refraction \citep{Lorimer2004hpa..book.....L}
\begin{equation}\label{eq:index_of_refraction_cold_plasma}
    n=\sqrt{1-\Bigg(\frac{f_{p}}{f}\Bigg)^{2}}.
\end{equation}
The plasma frequency is
\begin{equation}\label{eq: plasma frequency}
    f_{p}=\sqrt{\frac{e^{2}n_{e}}{\pi m_{e}}} \simeq 8.5 \, \textnormal{kHz} \, \left(\frac{n_{e}}{\textrm{cm}^{-3}}\right)^{1/2},
\end{equation}
where $e$ is the elementary charge, $n_{e}$ is the electron number density, and $m_{e}$ is the electron mass. The phase velocity is modified by $n$ as $v=c/n$ where $c$ is the speed of light in vacuum ($n=1$). In the interstellar medium (ISM), generally $f_{p} \ll f$ and so $n \approx 1-f_{p}^{2}/2f^{2}$. Since $n \lesssim 1$ within cold plasma, the phase velocity exceeds $c$.

In the early 80's \citet{Fiedler1987} measured the light curves of 36 extragalactic sources with the Green Bank interferometer at 2.7 and 8.1 GHz noting, ``unusual minima'' \citep{Fiedler1987} in these sources. The most dramatic dimming was seen in both the 2.7 GHz and 8.1 GHz light curves of quasar QSO (quasi-stellar object) 0954+658.

This \textit{extreme scattering event} (ESE) occurred over a period of approximately 80 days as opposed to more typical variations for this source, which occurred on the timescale of days. The ESE was characterized by its prolonged flux density minimum of approximately 60 days, which were flanked by two dramatic caustic spikes (focusing of light rays resulting in multiple imaging) at the beginning and end of the event each spanning approximately 15 days each. \cite{Fiedler1987} ruled out the possibility that this event was due to intrinsic variability since models based on an evolving emitter show flux density variations at a lower frequency lagging those at higher frequencies. However, this ESE showed simultaneous variations in 2.7 GHz and 8.1 GHz, atypical of its intrinsic flux variations which lag by 50 days between the two wavebands, which suggested it was due to an occulter of astronomical unit (AU) size at kiloparsec (kpc) distances. Variations and spikes in the light curve were thought to be caused by irregularities in ionised gas density and substructure. From the flux density profile of the ESE, assuming a spherical \textit{plasma lens} implied the occulter had electron densities $n_{e} \sim 4 \times 10^{4}\, \text{cm}^{-3}$, much larger than the diffuse ISM. These small plasma lenses are a type of \textit{tiny-scale ionised structure} (TSIS). Such small-scale structures in the ionised ISM may reach sizes as small as 70 - 100 km \citep[for example]{Rickett1977,Armstrong1995ApJ...443..209A}.

There have since been a handful of ESEs observed \citep{Stanimirovic2018}. It should be noted that the \textit{exact} definition of an ESE is unclear; the precise limiting characteristics of an ESE have not been mutually agreed upon between authors. Collectively however, it seems ESEs are characterized by their significant drop ($>50\%$) in radio ($\sim 1$ GHz) flux density over the course of weeks to months. Such dramatic flux density variation and long duration distinguishes itself from typical scintillation due to inhomogeneities of the electron density in the interstellar plasma which varies only a few percent over hours or days.

An early, well-known model of a plasma lens was explored by \cite{Clegg1998}. The \cite{Clegg1998} lens was modeled by a Gaussian-distributed electron column density. \cite{Clegg1998} found their model was able to produce some notable features of ESE light curves including a defocused region (the minima), and caustic spikes \citep[Figure 2]{Clegg1998}. The refractive nature of plasma lenses are described by Equation \ref{eq:index_of_refraction_cold_plasma} when the propagating wavelength is much smaller than the spatial variation of $n_{e}$. The radiation is refracted toward regions of lower phase speed and lower $n_{e}$. Multiple imaging was predicted by the model and later confirmed by an observation of QSO 2023+335 by \cite{Pushkarev2013A&A...555A..80P}. By considering the \cite{Fiedler1987} observations of QSO 0954+658, \cite{Clegg1998} found the plasma lens was 0.38 AU in size with electron density $10^{5} \, \textnormal{cm}^{-3}$. This corresponds to pressures of approximately $10^{6}-10^{9} \, \textnormal{K} \, \textnormal{cm}^{-3}$, ``\ldots well in excess of the average ISM pressure of roughly $4000 \, \textnormal{K} \, \textnormal{cm}^{-3}$'' \citep{Kulkarni1988gera.book...95K}. Such highly over-pressured structures suggest they should be transient in nature or would otherwise need to be situated within a high pressure environment.

Geometrical arguments have been proposed to explain the excessively large inferred density and pressures of plasma lenses. Not long after \citeauthor{Fiedler1987}'s (\citeyear{Fiedler1987}) ESE discovery, \cite{Romani1987} suggested ionised filamentary structures or plasma sheets as possible plasma lenses. If a cylindrical filament is viewed end-on, then a high column density is observed with a small transverse size. Thus, an illusion is created where if the observer assumes the object is spherical (as \citet{Clegg1998} did), then the density and inferred pressure would be greatly overestimated. Similarly, plasma sheets viewed edge-on or many plasma sheet layers could provide the high observed column densities. Such plasma sheets could exist in the ionization fronts or cooling instabilities of old supernova remnants where small scattering angles could occur through hundreds of face-on sheets, on average \citep{Romani1987}. The most extreme scattering, however, would occur due to corrugations or inhomogeneities in the sheets' structure \citep{Romani1987,Pen2014MNRAS.442.3338P}. Similarly, ionised shells of molecular clouds could provide the necessary scattering conditions \citep{Boldyrev2006ApJ...640..344B}. In the \cite{Boldyrev2006ApJ...640..344B} model, the scattering would be dominated by a few strong scattering events as opposed to the culmination of multiple smaller scattering events.

Unaware of extreme scattering by plasma lenses, \cite{Goldreich2006ApJ...640L.159G} proposed sheet-like structures primarily aligned along the line-of-sight were responsible for extreme scattering in the Galactic Centre. Motivated by \cite{Goldreich2006ApJ...640L.159G}, \citet{Pen2012MNRAS.421L.132P} model plasma lenses as under-dense current sheets ``\ldots as a generic solution to strong interstellar scattering'' including ESEs. Interestingly, these under-dense sheets act as convergent lenses as opposed to divergent. A real-time survey with the Australia Telescope Compact Array surveying $\sim 1000$ AGNs over $4-8$ GHz identified an ESE in the direction of PKS 1939-315 in 2014 \citep{Bannister2016}. A measurement of the electron column density as a function of time $N_{e}(t)$ for this event was achieved. Since the peak of the $N_{e}(t)$ profile corresponded with the minimum of the flux density profile, it was concluded that the ESE was caused by a divergent lens. Such evidence contradicts the \cite{Pen2012MNRAS.421L.132P} model.

ESEs are distinguished by their dramatic decrease in flux density with time. \cite{Lazio2000ApJ...534..706L} conducted multi-epoch Very Long Baseline Interferometry (VLBI) observations of PKS 1741-038 during an ESE. The source appeared to increase in angular size during the event. By the conservation of surface brightness of a purely refractive lens model, an increase in angular size should correspond to an increase in brightness, not the decrease that is characteristic of ESEs. \cite{Lazio2000ApJ...534..706L} explained their observations using both turbulent and refractive effects. \cite{Pen2012MNRAS.421L.132P} suggested that unresolved multiple images could explain the apparent increase in angular size.

Other ``extreme scattering'' phenomena such as intraday variability (IDV) of radio quasars, parabolic arcs in pulsar secondary spectra \citep{Stinebring2001ApJ...549L..97S}, and Galactic Centre scattering \citep{Goldreich2006ApJ...640L.159G} have been observed. All such phenomena are suspected to arise from the scattering of radiation by turbulent media, or plasma structures. All these phenomena are also plagued by the problem that the scattering plasma structures must be overpressured with respect to the diffuse ISM, to various degrees. Although it is suspected that turbulent media or plasma structures are responsible for the extreme scattering, it is unknown if they are indeed connected phenomena or not.

Only a handful of ESEs have been discovered even with extensive surveys \citep{Fiedler1994ApJ...430..581F,Lazio2001ApJS..136..265L,Pushkarev2013A&A...555A..80P}. Thus, ESEs suffer from small number statistics and their detection rate is inconclusive. This is further troubled by difficulties distinguishing between ``small'' and ``high'' amplitude variability. Short time-scale, small-amplitude ESEs are easily confused with interstellar scintillation, while long-timescale small-amplitude ESEs can be contaminated by intrinsic source variability \citep{Stanimirovic2018}. Automated methods such as the wavelet transformation of light curves by \cite{Lazio2001ApJS..136..265L} have been able to identify ESEs with moderate accuracy. While \cite{Lazio2001ApJS..136..265L} were able to identify 15 events from 12 sources, five ESEs previously found by eye were not identified using their wavelet transformation tool. Most ESE discoveries have been due to chance occurrence. \cite{Bannister2016}, however, were able to achieve a real-time ESE observation by monitoring $\sim 1000$ active galactic nuclei for changes in the continuum corresponding to the wavelength-squared dependence of the plasma refractive index indicating the beginning of a possible ESE. Dedicated monitoring for ESEs are needed to gather the necessary statistics and measurements that characterize ESEs.

Since \citeauthor{Clegg1998}'s (\citeyear{Clegg1998}) model, more complicated models have been developed to explain ESEs and the plasma lenses that are theorized to produce them. Whereas \citet{Clegg1998} considered a lens with an electron \textit{column} density distributed as a two-dimensional Gaussian on the sky, \citet{Er2018} considered power law, and exponentially distributed electron \textit{volume} densities. \citet{Er2018} found interesting behaviours in the magnification curves of these models including similar magnification curves between the well-studied Gaussian lens and their softened power-law lens. \citet{Rogers2019} generalized plasma lensing models to dual-component lenses using families of exponential and softened-power laws, finding more complicated magnification curves and caustic structures. Interestingly, \cite{Rogers2019} pointed out a degeneracy between two specific models -- a dual-component Gaussian lens, and a single spherically symmetric exponential lens. Since the index of refraction of plasma varies with the wavelength of the signal, this degeneracy may be broken by the active observation at various wavelengths of an ESE \citep{Rogers2019}. \cite{Er2019} further expanded the library of analytical plasma lens models to elliptical lenses following exponential and softened power law distributions.

Extending beyond spherical and elliptical lenses, \cite{Rogers2020} explored toy filamentary models confined by helical magnetic fields. The cylindrical geometry, when viewed along the long axis, may create the illusion of a small transverse size without the issue of being over-pressured. \cite{Rogers2020} were able to reproduce the magnification of spherical lenses even with the cylindrical geometry, but determined that these models were physically unrealistic from the resulting dispersion measure.

More recently, \cite{Er2022MNRAS.509.5872E} investigated plasma lensing as modeled by a double-plane, each with a projected Gaussian-distributed electron density. Such an approach is akin to scattering by a pair of effective planes as opposed to volumetric scattering. Effective scattering planes are often used to describe the effects of gravitational lensing \citep{Schneider1992grle.book.....S}, and scintillation \citep[for example]{Hill2005ApJ...619L.171H,Reardon2020ApJ...904..104R,Sprenger2022MNRAS.515.6198S,Stinebring2001ApJ...549L..97S}. While some image properties from a single Gaussian lens mimicked those produced by a double-lens, there were some properties which distinguished between the two scenarios. In particular, the time delay (and time delay-frequency relation) between the two models were sufficiently different.

\subsection{Problems with Plasma Lens Models}
There are problems concerning the properties of the plasma structures that are predicted to produce ESEs, which speak more broadly to the physics of the ISM. The primary and most accessible problem to be addressed is to better constrain the geometries and physical characteristics of the plasma lenses that produce ESEs. Consequently, additional questions may be addressed by relating the physical properties of plasma lenses to the problem of over-dense, over-pressured models. Given the relatively small scale of these plasma structures and their predicted high densities and pressures, it is unknown how prevalent these objects are, what the origins of these plasma structures are, and how they might influence the physics of the ISM. ESEs provide an opportunity ot study dense structures in the ISM at small length scales that woul be difficult to access otherwise. A plausible model of ESEs would provide new insights in the the physics of the ISM.

One primary concern with plasma lenses is how they are predicted to be highly overpressured relative to the surrounding ISM. Without some confinement mechanism, their high pressures suggest that plasma lenses must be transient structures. Models based on thin sheets viewed (nearly) edge-on could provide a means to reduce the lens pressure. It is possible that other geometries and methods of confinement could provide the necessary conditions to produce the lenses required for ESEs \citep[for example]{Romani1987,Rogers2020}. Between theory and observations, nothing is conclusive at this point concerning the geometric arguments made to address the over-pressure problem.

A big mystery concerns the origin of ESEs and the plasma lensing structures that produce them. Whether these ESEs originate from ionised turbulence or from discrete lensing structures is unknown. Supernova remnant (SNR) shells, bubble walls, cloud edges, and shocks have been postulated as sites of ESEs \citep{Romani1987}. \cite{Hamidouche2007A&A...468..193H} support the idea that ESEs can be produced by turbulence, however the few instances of multiple-imaging detected during an ESE suggest they are produced by discrete plasma lenses. A turbulent origin of ESEs has also been suggested, but volumetric turbulence, or \textit{volumetric ray-tracing} more generally, has yet to be explored in investigating ESEs \citep{Stanimirovic2018}.

Investigating ESEs and plasma lenses will provide significant insights into the physical processes of the ISM. These small structures are on the scale of turbulent dissipation, and their properties may yield clues informing poorly understood physical processes at these scales. These overpressured plasma structures could be a source of significant heating in the ISM depending on their volume filling factor. A measurement of the volume filling factor, however, has not yet been conducted. \cite{Stanimirovic2018} suggest tiny scale ionised structures (TSIS) such as plasma lenses are a ``\ldots unique probe of stellar feedback efficiency since they are likely produced by local turbulence enhancements in places like SNRs, stellar bubbles, and winds'' and ``\ldots deserve a major spotlight and attention.'' 

\subsection{Method Outline}

Our aim is to construct ESE light curves using numerical volumetric ray-tracing methods. Rays are traced between the source (such as a pulsar) and the observer through a lensing medium. The geometry of the system is set up similarly to a gravitational lensing scenario such as \citet[Figure 5]{Narayan1996}. The source is placed a distance $D_{s}$ away from the observer, the lens' origin is placed a distance $D_{d}$ from the observer, and the distance between the lens and the source is given by $D_{ds}=D_{s}-D_{d}$. In fact, the problem itself draws many analogies with gravitational lensing, which we use as a test of our method (Section \ref{sec:TestCases}).

Ray-tracing is calculated by solving the \textit{eikonal equation} through a medium of spatially varying refractive index (Section \ref{sec:Ray-tracing}). The eikonal method requires that a model specifying $n(\vec{x})$ is provided. Multiple-imaging is accounted for by mapping pixels on the observer plane to overlapping or clustered points on the source plane. The mapping of these pixels is done by propagating a wave front formed by Delaunay triangulation over rays traced at the triangle vertices (Section \ref{sec:Wavefront_Propagation_with_Delaunay_Triangulation}). Consequently, magnification maps (Section \ref{sec:Magnification Maps}), intensity maps (Section \ref{sec:Intensity_Maps}), and light curves (Section \ref{sec:Light_Curves}) are constructed from mapping the brightness of a source to calculate the intensity by the observer/detector. Whereas much of the literature has modeled plasma lenses and the corresponding light curves analytically, numerical ray-tracing methods offer an opportunity to explore new lens models, especially those defined by volumetrically extended structures such as regions of plasma turbulence. By approaching the investigation numerically, solutions can be made much more general. Instead of projecting scattering effects onto an effective two-dimensional scattering plane, we use full three-dimensional lens models.

\section{Ray-Tracing}
\label{sec:Ray-tracing}
Snell's Law adequately describes refraction at the interface between two uniform media. More realistically, when the medium that light passes through varies continuously in refractive index, Snell's Law becomes inadequate to describe the trajectory of light through the medium. In regions where the wavelength of light is small compared to the size of the optical system, the amplitude and direction of propagation of a wave does not change considerably for several wavelengths and can be approximated as a plane wave. General solutions of the wave equation are described with a frequency $\omega$ and wave vector $k=\omega/c$ \citep{Romer2006theoretical}
\begin{equation}
    u_{k}(t,\vec{x}) = \varphi_{0}\mathrm{e}^{-i(\omega t - \vec{k} \cdot \vec{x})} = \varphi_{0} \mathrm{e}^{-ik(ct - \vec{n} \cdot \vec{x})} = \varphi_{k}(\vec{x})\mathrm{e}^{ik[S(\vec{x})-ct]}
\end{equation}
where $\varphi_{k}(\vec{x})$ is a real number quantifying the wave amplitude, and $S(\vec{x})$ is a real function known as the \textit{eikonal function}. The eikonal function determines the phase of the wave at position $\vec{x}$ such that $S(\vec{x})=\textrm{constant}$ represents surfaces of constant phase. Ray-tracing is performed by solving the \textit{eikonal equation}
\begin{equation}\label{eq: eikonal equation solved Romer 2006 Eq 7.31}
    \frac{\mathrm{d}}{\mathrm{d}s}\left(n\frac{\mathrm{d}\vec{x}}{\mathrm{d}s}\right)=\vec{\nabla}n,
\end{equation}
where the index of refraction $n(\vec{x})$ spatially varies, $\vec{x}$ are the ray trajectories, and the arc length $s$ serves as the curve parameter. The eikonal equation describes families of rays with trajectories orthogonal to the level surfaces of constant phase and is solved numerically by re-parameterizing Equation \ref{eq: eikonal equation solved Romer 2006 Eq 7.31} in terms of time $t$. To begin, we define a parameter
\begin{equation}\label{eq: q=n*dx/ds definition eikonal parameterization}
    \vec{q} \equiv n \frac{\mathrm{d}\vec{x}}{\mathrm{d}s}  = n\frac{\mathrm{d}\vec{x}/\mathrm{d}t}{\mathrm{d}s/\mathrm{d}t} = n \frac{\vec{v}}{\Vert \vec{v} \Vert},
\end{equation}
\begin{subequations}\label{eq: parameterized eikonal}
where $\vec{v}$ is the ray velocity and $v$ is the corresponding speed in the medium is given by
\begin{equation} \label{eq: ds/dt=c/n eikonal parameterization}
    v = \Vert \vec{v} \Vert = \frac{\mathrm{d}s}{\mathrm{d}t} = \frac{c}{n}.
\end{equation}
With $\vec{q}$ and $\mathrm{d}s/\mathrm{d}t$, Equation \ref{eq: eikonal equation solved Romer 2006 Eq 7.31} is rewritten as
\begin{equation} \label{eq: dq/dt = c/n nabla n eikonal parameterization}
    \frac{\mathrm{d}\vec{q}}{\mathrm{d}t} = \frac{c}{n}\vec{\nabla}n.
\end{equation}
Using the definition of $\vec{q}$ (Equation \ref{eq: q=n*dx/ds definition eikonal parameterization}) and $\mathrm{d}s/\mathrm{d}t$ (Equation \ref{eq: ds/dt=c/n eikonal parameterization}), the velocity of the ray trajectories are
\begin{equation}\label{eq: dx/dt = c/n^2 q eikonal parameterization}
    \frac{\mathrm{d}\vec{x}}{\mathrm{d}t} = \frac{c}{n^{2}} \vec{q}.
\end{equation}
\end{subequations}
Equations \ref{eq: ds/dt=c/n eikonal parameterization} to \ref{eq: dx/dt = c/n^2 q eikonal parameterization} are solved numerically for the ray trajectories using the initial conditions for the position $\vec{x}_{0}$ and $\vec{q}_{0}=n \vec{v}_{0}/\Vert \vec{v} \Vert$ (Equation \ref{eq: q=n*dx/ds definition eikonal parameterization}).

\subsection{Numerically Solving for the Ray Trajectories}
\label{sec:Numerically_Solving_for_the_Ray_Trajectories}
Computing ray trajectories requires integrating the parameterized eikonal equation (Equation \ref{eq: parameterized eikonal}). A custom fifth-order Runge-Kutta (RK5) integrator based on the Dormand-Prince method \citep{DORMAND198019,Press2007numerical} is written and tested to numerically solve the parameterized system. A custom RK5 integrator is primarily advantageous for tracing many rays simultaneously and managing stopping conditions.

\begin{figure}
\plotone{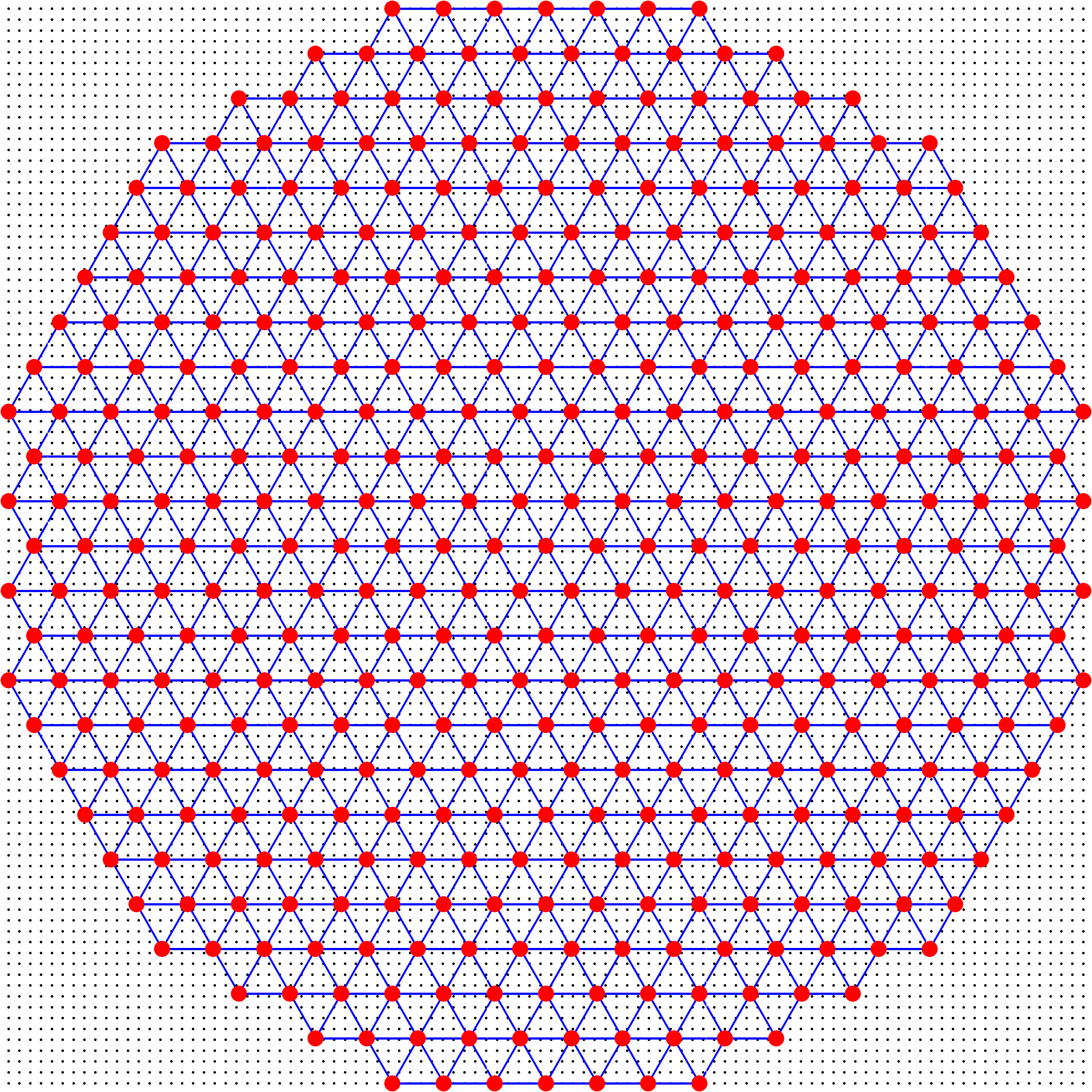}
\caption{Illustrative figure of initial ray positions (large red dots) on the observer plane distributed in a grid of equilateral triangles (Section \ref{sec:Numerically_Solving_for_the_Ray_Trajectories}). Delaunay triangulation (blue lines) are formed from connecting the initial ray positions, which act as the triangle vertices (Section \ref{sec:Wavefront_Propagation_with_Delaunay_Triangulation}). A grid of pixels (small black dots) are identified with the Delaunay triangle they reside within (Section \ref{sec:Accounting_for_Multiple_Imaging}). \label{fig:imgplane_px_fullgrid_DT_initalRays.png}}
\end{figure}

Rays are initially placed on the image or observer plane (the terms ``image'' and ``observer'' planes will be used interchangeably) such that each ray forms the vertices of a pattern of equilateral triangles (Figure \ref{fig:imgplane_px_fullgrid_DT_initalRays.png}).
From a distant source such as a pulsar or quasar, the rays arrives at the observer (Earth) with a small angle. If gravitational lensing and Einstein rings are indicative of the angle of light rays detected, it suffices to launch the initial rays with an angle of order $\sim \theta_{E}$ (the Einstein angle) primarily in the normal (+z) direction from the observer (x-y) plane toward the source plane. For our volumetric ray-tracing method with plasma lenses, $\theta_{E}$ serves as a starting point to launch the initial rays, which is then fine-tuned. Although perhaps it is counter-intuitive to trace rays from the observer plane to the source plane, this is necessary to account for multiple-imaging (see Section \ref{sec:Wavefront_Propagation_with_Delaunay_Triangulation}). Once the rays are launched from the initial positions, they are propogated through a medium with spatially varying continuous index of refraction by solving the eikonal equation with the RK5 integrator. Several models of plasma structures with spatially-varying refractive index are discussed in Section \ref{sec:Plasma Lens Examples}. The integrator is halted once the rays reach the source plane. Parameters such as $\vec{x}$, $\vec{q}$, and $s$ on the source plane are obtained by  interpolation of the ray one time step before and after the ray crosses the source plane.

A map is created from ray-tracing since both ray positions on the observer plane and the corresponding position on the source plane are known. This mapping is the cornerstone of this investigative technique for exploring plasma lenses and ESEs because it links the brightness distribution between the source and the observer regardless of how the medium is structured. It allows for the mapping of pixels of a detector (the observer plane) to the brightness from a distance source (Section \ref{sec:Intensity_Maps}) and the construction of light curves (Section \ref{sec:Light_Curves}).

\section{Wavefront Propagation with Delaunay Triangulation}
\label{sec:Wavefront_Propagation_with_Delaunay_Triangulation}
Ray trajectories are directed in the normal direction with respect to the wavefronts. By calculating the ray trajectories through a refracting medium, wavefront elements are approximated by Delaunay triangles. For a two-dimensional set of points, the Delaunay triangulation criterion requires that the circumscribed circle or circumcircle of each triangle contains no other points in its interior. Consequently, the formation of Delaunay triangles is such that triangles with larger minimum angles are formed with preference over those with smaller minimum angles. It is worth noting that the generation of Delaunay triangles can be degenerate for a given set of points, but this is inconsequential as long as the scheme covers the triangulated area.

MATLAB's Delaunay functions create a Delaunay object that contains the points or vertices of each Delaunay triangle, and also contains a ``connectivity list'' establishing the connection between the three vertices forming a Delaunay triangle. Points are listed in an array such that each row is a point whose $x,y,z$ coordinates are listed by column. The connectivity list contains the row numbers of the three vertices of each triangle, which point to the coordinates of each vertex, and is established at the initial positions of the ray trajectories (Figure \ref{fig:imgplane_px_fullgrid_DT_initalRays.png}). Once generated on the image plane, the connectivity list provides a static list of connections between triangle vertices defined by rays. Thus, this connectivity list is preserved throughout the rays' trajectories, effectively tracking the propagation of the wavefront.

\subsection{Accounting for Multiple Imaging}
\label{sec:Accounting_for_Multiple_Imaging}
Delaunay triangles provide a significant advantage for handling multiple imaging. Recall that this technique begins with rays on the observer plane which are ray-traced through a medium with spatially varying refractive index to intersect the source plane. Delaunay triangles on the source plane may be distorted from their initial positions on an equilateral grid on he observer plane. Each Delaunay triangle is identified and tracked thanks to the connectivity list.

Accounting for multiple imaging involves creating a mapping between pixels on the observer plane and the positions where these pixels end up as points on the source plane after travelling through the medium. Pixel centres are first created on the observer plane in a grid formation. Each pixel is then identified with the Delaunay triangle it resides within. Assuming barycentric coordinates are preserved as the wavefront of Delaunay triangles propagates through the medium, then the pixels are mapped onto the source plane since the connectivity list of Delaunay triangles between the observer plane and source plane are known (Section \ref{sec:Ray-tracing}).

\begin{figure}
\plotone{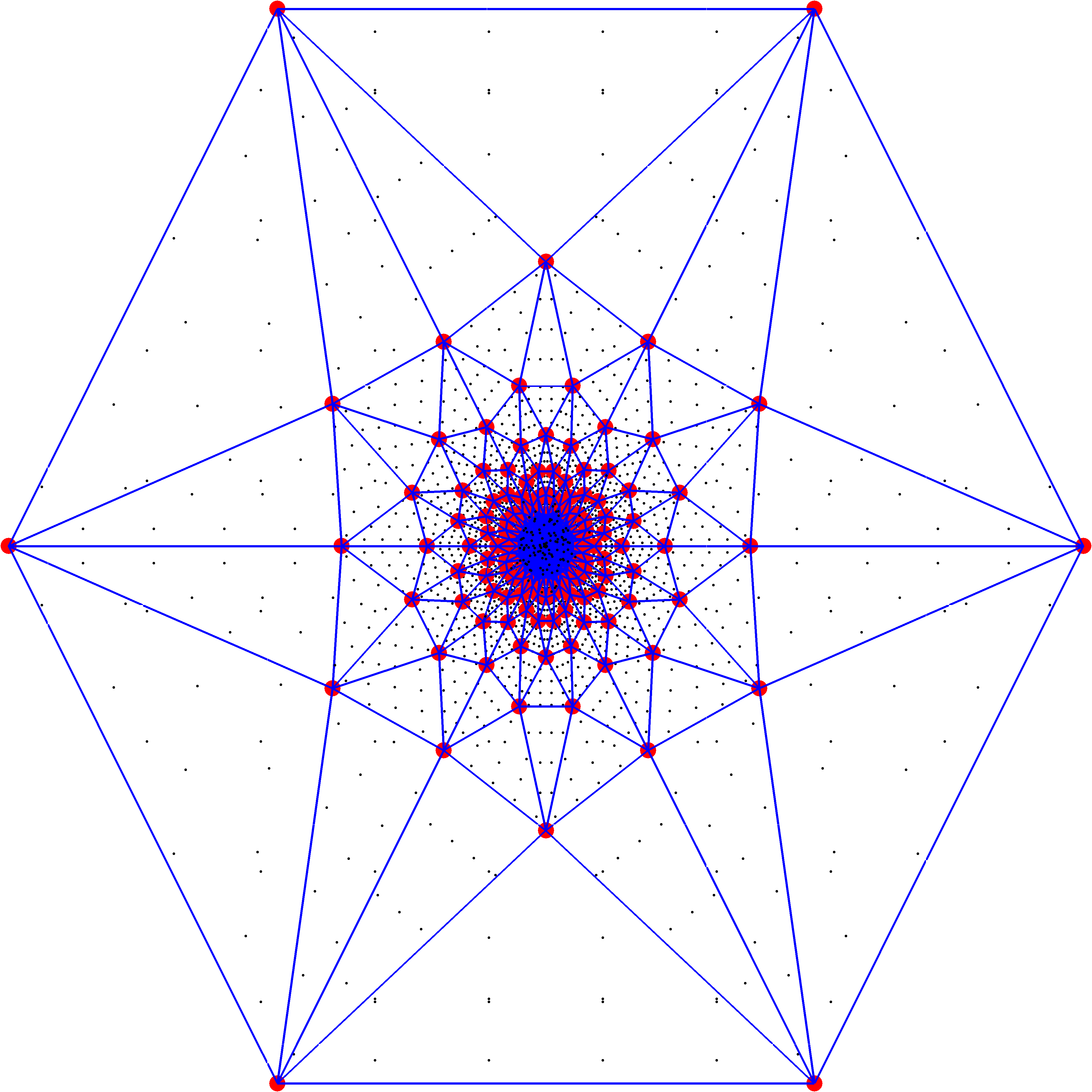}
\caption{Illustrative figure of lensed ray positions (large red dots) on the source plane (Section \ref{sec:Numerically_Solving_for_the_Ray_Trajectories}) corresponding to Figure \ref{fig:imgplane_px_fullgrid_DT_initalRays.png}. The connectivity list tracks the Delaunay triangle wavefront (blue lines) from the observer plane to the source plane (Section \ref{sec:Wavefront_Propagation_with_Delaunay_Triangulation}). Pixels on the observer plane are mapped onto the source plance as points (small black dots) because their barycentric coordinates are preserved (Section \ref{sec:Accounting_for_Multiple_Imaging}). A cluster of points, Delaunay triangles, and rays in the centre of the image are too densely spaced to be individually distinguished. Multiple-imaging is conducive in the centre of the figure where the points are most tightly clustered because the brightness within that small region will map to many pixels on the observer plane. This specific figure uses a point mass gravitational lens (Section \ref{sec:GravitationalPointMassLens}) for illustrative purposes only. \label{fig:srcplane_px_fullgrid_DT_initalRays.png}}
\end{figure}

Some of the points may be closely clustered together or even overlap on the source plane. Consequently, this mapping accounts for multiple-imaging as the brightness of many pixels on the observer plane map to the same point or cluster of points in close proximity on the source plane (Figure \ref{fig:srcplane_px_fullgrid_DT_initalRays.png}). This mapping is used to generate magnification maps (Section \ref{sec:Magnification Maps}) and intensity maps (Section \ref{sec:Intensity_Maps}).

\section{Magnification Maps}
\label{sec:Magnification Maps}
Delaunay triangles provide an intuitive way to understand and calculate magnification. Recall that the Delaunay triangles are formed from connecting the vertices or rays that propagate through a medium with spatially varying index of refraction. The rays may scatter and take non-uniform, irregular trajectories through the medium, which deforms the wavefront or surface of Delaunay triangles. The magnification in a triangle is given simply by
\begin{equation}\label{eq:magnification_Narayan}
    \mu=\frac{A_{i}}{A_{s}},
\end{equation}
where $A_{i}$ and $A_{s}$ are the respective areas covered by a mapped triangle on the image plane and on the source plane. The magnification is assigned to the centroid of each triangle on the observer plane and interpolated onto the observer plane pixels to generate a magnification map. As no further calculations are performed from the magnification map, one magnification map is shown in Section \ref{sec:GaussianLens} for brevity.

\section{Intensity Maps}
\label{sec:Intensity_Maps}
Intensity maps provide a useful way to visualize the images created by this method of ray-tracing and wavefront propagation, and are a precursor to constructing light curves (Section \ref{sec:Light_Curves}). Mapping pixels on the observer plane to points on the source plane (Section \ref{sec:Accounting_for_Multiple_Imaging}) is key to constructing intensity maps.

First the brightness distribution of the source is constructed on the source plane. The brightness profile of the source is typically assumed to be Gaussian distributed
\begin{equation}\label{eq:gaussian source brightness}
    f(\vec{x})=\exp{\left[ -\frac{(\vec{x}-\vec{x}_{s})^{2}}{2\sigma^{2}}\right]},
\end{equation}
where $\sigma^{2}$ is the variance ($\sigma$ is the standard deviation) and $\vec{x}_{s}$ is the centre of the source, although any brightness distribution may be used. An intensity map on the source plane is produced by first creating a grid of pixel centres on the source plane and then calculating the brightness at each pixel centre based on the source's position.

Having established the mapping of points on the source plane to pixels on the observer plane, the brightness of the points on the source plane is mapped directly onto the pixels of the observer plane. Note that this mapping does not directly use the source plane pixels; rather, the image plane pixels are mapped back to coordinates on the source plane, allowing us to assign intensity values from the source model to the image plane pixels. This procedure generates an intensity map of the image, which automatically accounts for multiple imaging since multiple pixels on the observer plane may sample the same points on the source plane. Instances of multiple imaging are clear especially in the examples and test cases such as gravitational lensing by a point mass presented in Section \ref{sec:TestCases}. 

\subsection{Subpixels}
\label{sec:Subpixels}
Creating intensity maps from mapping pixels to points on the source plane is an effective way to visualize images resulting from lensing systems. However, to decrease noise in the intensity map, each pixel of the observer plane is subdivided into subpixels to calculate an averaged intensity value. In this work, the subdivision of pixels is controlled by an \textit{oversampling parameter ($OSP$)} such that pixels are divided into $OSP \times OSP$ subpixels. Typically $OSP=2$ is chosen as a good compromise between image fidelity and computation speed. The same mapping of pixels from Section \ref{sec:Accounting_for_Multiple_Imaging} is used to map the subpixels' barycentric coordinates from the observer plane to the source plane, the source brightness is calculated for each point (mapped from a subpixel), and the brightness is mapped back to the observer plane to construct the intensity map. The difference is, however, that the intensity of a pixel is calculated as the average over its subpixels. With mapping the subpixels to the source plane, the region where the points are on the source plane is more thoroughly sampled. Consequently, the intensity averaged from the subpixels is a more accurate reflection of the brightness of the region sampled compared to using single pixel centres.

\section{Light Curves}
\label{sec:Light_Curves}
Time-dependent changes in the lensing system may cause changes in the total intensity measured, resulting in a light curve. We assume that the lensing medium and observer are stationary while the source object moves, and we construct a light curve from intensity maps generated for a sequence of time steps. Telescope beam effects can be (optionally) introduced in this step by assuming a beam comprised of spatially variable sensitivity on the observer plane. We typically assume a Gaussian beam sensitivity profile or a flat beam sensitivity, which is the default.

To calculate the total intensity from the intensity maps (Section \ref{sec:Intensity_Maps}), the intensity of each pixel in the map is multiplied by the area of the pixel, and summed for all pixels in the map. This process is repeated for a sequence of times steps to generate a light curve. The light curve is normalized by the total intensity on the image plane when the source object is far off-axis with the lens and the lensing effects are negligible.

\section{Test Cases}\label{sec:TestCases}
We test our ray-tracing method by comparing to a simple point mass gravitational lens solution, and diverging point lens solution, both with known analytic solutions.

\textbf{For numerical calculations, it is conventional to compute in dimensionless or scaled quantities, which we denote with an overscript tilde.}

\subsection{Gravitational Point Mass Lens}
\label{sec:GravitationalPointMassLens}
Gravitational lensing by a point mass, and gravitational lensing more generally, is well-studied \citep[for example]{Schramm1987A&A...174..361S,Kayser1988A&A...191...39K,Schneider1992grle.book.....S,Narayan1992_10.2307/53917,Narayan1996} and there exists an especially simple analytical solution for lensing by a point mass. As light propagates from a source, its trajectory follows geodesic paths in the curvature of spacetime, which are observed as curved paths. For small perturbations by a gravitational potential $\Phi$ in locally flat Minkowski spacetime, the curvature of light by a gravitational lens is described by an \textit{effective index of refraction} \citep{Schneider1992grle.book.....S,Narayan1996}
\begin{equation}
    n=1-\frac{2}{c^{2}}\Phi=1+\frac{2}{c^{2}}|\Phi|.
\end{equation}
For a gravitational lens that is a point mass $M$, the gravitational potential is given by
\begin{equation}
    \Phi=-\frac{GM}{r}.
\end{equation}
Consequently, the effective index of refraction for a point mass is
\begin{equation}
    n=1+\frac{2}{c^{2}}\frac{GM}{r},
\end{equation}
which can be simplified by collecting the constant terms such that
\begin{equation}\label{eq:gravitationalLens_pointMass_n_dimensional}
    n=1+\frac{k}{r}
\end{equation}
where $k=2GM/c^{2}$ is the corresponding Schwarzschild radius. The Schwarzchild radius provides a natural length scale such that the effective index of refraction (Equation \ref{eq:gravitationalLens_pointMass_n_dimensional}) in terms of the dimensionless radius $\tilde{r}=r/r_{0}$, where $r_{0}$ is the length scale ($r_{0}=k$ in this case), is written as
\begin{equation}\label{eq:gravitationalLens_pointmass_n_dimensionless}
    n=1+\frac{1}{\tilde{r}}.
\end{equation}
The corresponding dimensionless gradient is
\begin{equation}
    \vec{\tilde{\nabla}}n = - \frac{1}{\tilde{r}^{2}} \hat{r}.
\end{equation}
With $n$ and $\vec{\tilde{\nabla}}n$ modeled for the point mass lens, the eikonal equation (Equation \ref{eq: eikonal equation solved Romer 2006 Eq 7.31}) is solved numerically for the ray trajectories (Section \ref{sec:Ray-tracing}), followed by the Delaunay triangle wavefront (Section \ref{sec:Wavefront_Propagation_with_Delaunay_Triangulation}), which results in the magnification map (Section \ref{sec:Magnification Maps}), intensity map (Section \ref{sec:Intensity_Maps}), and light curve (Section \ref{sec:Light_Curves}).

Einstein rings are a visually striking consequence of multiple-imaging due to gravitational lensing by a point mass $M$. When a source lies directly behind the lens along the line of sight, it is imaged as an Einstein ring of characteristic \textit{Einstein angle}
\begin{equation}\label{eq:theta_E einstein radius NB96 Eq21 dimensional}
    \theta_{E} = \sqrt{\frac{4GM}{c^{2}}\frac{\Dds}{\Dd\Ds}},
\end{equation}
which is a well-known analytic and observed result \citep[for example]{Schneider1992grle.book.....S,Narayan1996}. Scaling $\theta_{E}$ by the Schwarzchild radius yields
\begin{equation}\label{eq:theta_E einstein radius NB96 Eq21 dimensionless}
    \theta_{E} = \sqrt{2 \frac{\tilde{D}_{ds}}{\tilde{D}_{d}\tilde{D}_{s}}}.
\end{equation}

\begin{figure}
\plotone{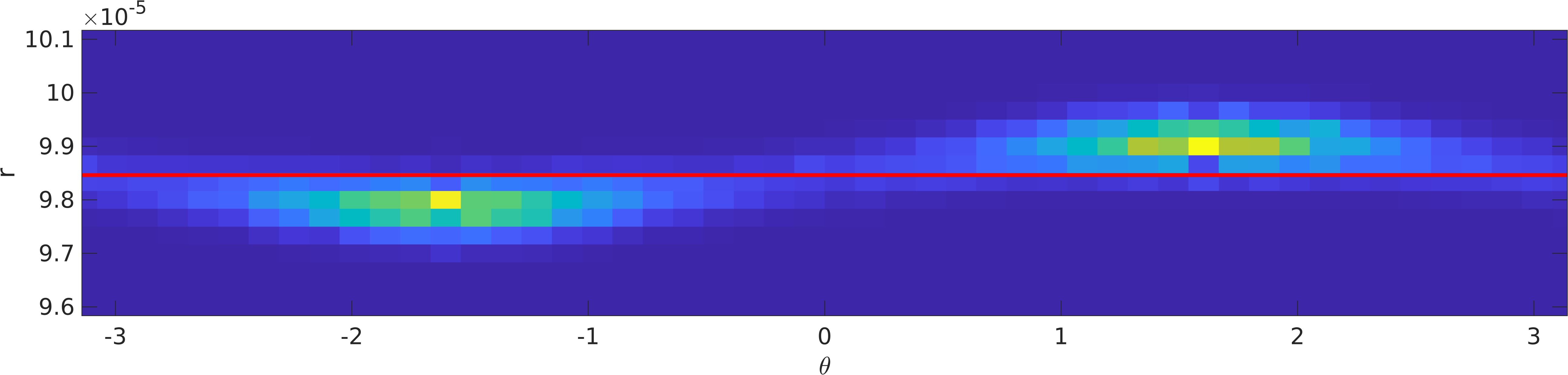}
\caption{Testing the validity of the ray-tracing method by comparing the numerical results with analytic results of an Einstein ring. $r$ is the angular radius of the Einstein ring and $\theta$ is the angle around it. The numerically ray-traced pixels from the source plane are mapped to an Einstein ring (multi-coloured) on the observer plane. A red line representing the analytic Einstein ring (Equation \ref{eq:theta_E einstein radius NB96 Eq21 dimensionless}) is overlaid on top of the numerical result. The two results match on the order of $\sim 10^{-6}$, which supports the validity of the numerical method. Parameters were chosen such that $k=1 \, \textrm{AU}$ and $D_{s} = 1 \, \textrm{kpc}$, which are sizes representative of pulsar ESEs.}
\label{fig:compare numeric to analytic point mass theta_E ring_plot}
\end{figure}

To test the validity of the ray-tracing procedure described previously, the resultant size of the Einstein ring on the observer plane intensity map must agree with the analytic Einstein ring size. Figure \ref{fig:compare numeric to analytic point mass theta_E ring_plot} demonstrates that the numerically ray-traced pixels on the source plane map to an Einstein ring of the same size as the analytic Einstein ring as calculated by Equation \ref{eq:theta_E einstein radius NB96 Eq21 dimensionless}, which supports the validity of the numerical ray-tracing method. 

\begin{figure}
\plotone{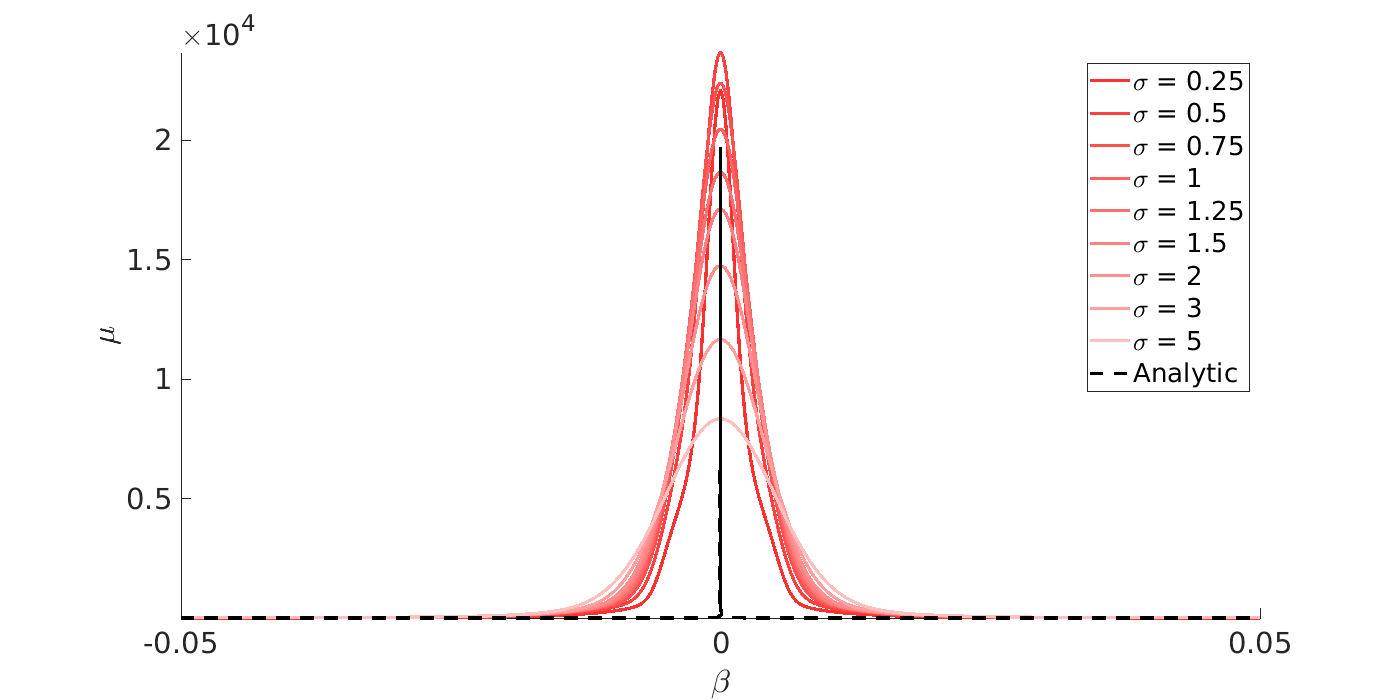}
\caption{Convergence of the numerical ray-traced solution as the Gaussian source brightness becomes more point-like (the standard deviation $\sigma$ (Equation \ref{eq:gaussian source brightness}) becomes smaller, shades of red lines) to the analytic solution (black dashed line) for a point mass gravitational lens. Total intensity or magnification $\mu$ is normalized, and is plotted as a function of angle on the source plane $\beta$. The gravitational point mass and source are aligned at $\beta=0$. A slight bulging effect is noticeable in the $\sigma=0.25$ light curve due to the small source size moving moving across regions with a higher density of points on the source plane. \label{fig:compare2analytic_pointmass}}
\end{figure}

Furthermore, the light curves computed numerically by ray-tracing should agree with the analytic result for a point mass gravitational lens \citep{Narayan1996}
\begin{equation}\label{eq: mu total magnification point mass}
    \mu=\frac{y^{2}+2}{y\sqrt{y^{2}+4}}
\end{equation}
where $y=\beta/\theta_{E}$ and $\beta$ is the angle on the source plane with respect to the optic axis of the lensing system \citep[Figure 5]{Narayan1996}. Figure \ref{fig:compare2analytic_pointmass} shows the normalized light curves for a point mass lens. Red lines of different shades in Figure \ref{fig:compare2analytic_pointmass} represent different sizes of the source brightness, which are compared with the analytic solution for a point source (black dashed line). As the Gaussian source becomes more point-like, the solution converges to the analytic solution, supporting the validity of this method.

\subsection{Point Diverging Lens}
\label{sec:PointDivergingLens}
As a toy model, the point diverging lens serves as a useful test case to compare the numerical ray-tracing with another analytic result \citep{Safonova2001PhRvD..65b3001S}. In general, the index of refraction for the point diverging lens is described by
\begin{equation}\label{eq:point diverging lens dimensional}
    n=1-\frac{k}{r},
\end{equation}
which is written in dimensionless form as
\begin{equation}\label{eq:point diverging lens dimensionless}
    n=1-\frac{1}{\tilde{r}}
\end{equation}
where $0 \leq k \leq 1$. Comparing the point diverging lens (Equation \ref{eq:point diverging lens dimensional}) to the point mass gravitational lens (Equation \ref{eq:gravitationalLens_pointMass_n_dimensional}), it is easy to notice that the two equations are different only by a change in sign of the second term resulting in a diverging lens. 

\begin{figure}
\plotone{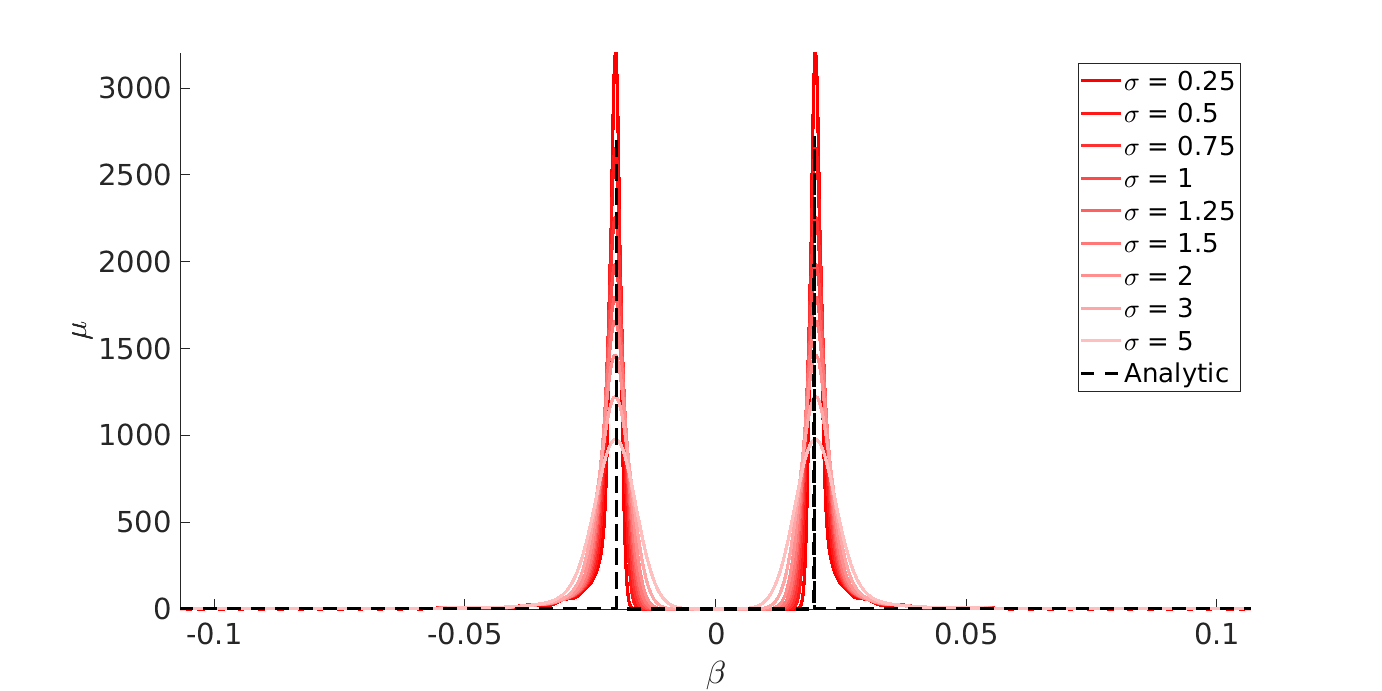}
\caption{Convergence of the numerical ray-traced solution as the Gaussian source brightness becomes more point-like (the standard deviation $\sigma$ (Equation \ref{eq:gaussian source brightness}) becomes smaller, shades of red lines) to the analytic solution (black dashed line) for a divergent point lens. Total intensity or magnification $\mu$ is normalized, and is plotted as a function of angle on the source plane $\beta$. The divergent point lens and source are aligned at $\beta=0$. A slight ringing effect is noticeable in the $\sigma=0.25$ and $\sigma=0.5$ light curves due to the small source size moving in gaps between points on the source plane. \label{fig:compare2analytic_divergentpointmass_N1e3_Ds0.1pc_Nx200}}
\end{figure}

Similar to the gravitational point mass lens (Section \ref{sec:GravitationalPointMassLens}), the light curves produced numerically by ray-tracing should converge to the analytic light curve for a point source. \cite{Safonova2001PhRvD..65b3001S} found the toy divergent point lens model produced a light curve described by 
\begin{equation}\label{eq: mu total magnitifcation divergent point lens Safonova Eq.35}
    \mu = \frac{y^{2}-2}{y\sqrt{y^{2}-4}}.
\end{equation}
Figure \ref{fig:compare2analytic_divergentpointmass_N1e3_Ds0.1pc_Nx200} demonstrates the convergence of the numeric solution as the size of the source becomes point-like (red lines) to the analytic solution (black dashed line). The convergence of the numeric light curve to the analytic light curve further supports the validity of the numerical ray-tracing method.

\section{Plasma Lens Examples}
\label{sec:Plasma Lens Examples}
Having tested our volumetric numerical ray-tracing method against the analytic cases of Section \ref{sec:TestCases}, we now use volumetric plasma lens models. Specifially, we will use a \cite{Clegg1998}-inspired Gaussian lens and an analytic model of a non-self-gravitating, magnetized, rotating filament \citep{Grafton2023MNRAS.522.1575G}.

In the most general case, the model for a plasma lens describes the particle mass density $\rho$. Assuming most of particles are hydrogen $\rho \approx \rho_{H}$, the hydrogen mass density $\rho_{H}$ is converted to an electron number density via
\begin{equation}\label{eq:electron volume density ne=rhoH*X/mu*mH}
    n_{e}=\frac{\rho_{H}X}{\mu m_{H}},
\end{equation}
where $X$ is the ionisation fraction, $\mu$ is the mean molecular mass, and $m_{H}$ is the mass of a hydrogen atom. For electromagnetic radiation at frequency $f$ propagating through a cold, ionised plasma such as the ISM, the electron number density contributes to the calculation of the plasma frequency $f_{p}$ (Equation \ref{eq: plasma frequency}), which in turn determines the index of refraction $n(\vec{x})$ (Equation \ref{eq:index_of_refraction_cold_plasma}). The refractive index and its gradient $\vec{\nabla}n$ are the main components needed to solve eikonal equation (Equation \ref{eq: parameterized eikonal}) and calculate ray trajectories (Section \ref{sec:Ray-tracing}). The signal frequency $f=10^{9} \, \textrm{Hz}$ is adopted as ESEs are detected in the radio ($\sim \textrm{GHz}$).

\subsection{Gaussian Lens}
\label{sec:GaussianLens}
\cite{Clegg1998} presented a classic explanation for the phenomenon of ESEs. In their model of a plasma lens, \cite{Clegg1998} considered a Gaussian-distributed electron column density profile, which produced the notable caustic spikes and flux drop that are characteristic of ESEs. Here, the \cite{Clegg1998} plasma lens model serves as inspiration for a full three-dimensional plasma lens with Gaussian-distributed electron density.

In dimensionless form the electron volume density for the \cite{Clegg1998}-inspired spherical Gaussian plasma lens is
\begin{equation}
    n_{e} = n_{e0}\exp{\left( - \frac{\tilde{r}^{2}}{2} \right)},
\end{equation}
where the scaling of the spherical radius follows $r = r_{0}\tilde{r} = \sigma \tilde{r}$, $\sigma$ is the standard deviation, and $n_{e0}=\rho_{0}X/\mu m_{H}$ is the electron volume density at the core $r=0$. The electron volume density is used to calculate the plasma frequency $f_{p}$ (Equation \ref{eq: plasma frequency}), which is used to calculate the refractive index $n$ (Equation \ref{eq:index_of_refraction_cold_plasma}). Since the Gaussian lens possesses spherical symmetry, the dimensionless gradient of the refractive index is
\begin{equation}\label{eq: gaussian lens dimensionless gradient of the index of refraction cold plasma}
    \vec{\tilde{\nabla}}n = \frac{1}{2\sqrt{1-\left(f_{p}/f\right)^{2}}}\left(\frac{f_{p}}{f}\right)^{2}\tilde{r}\;\hat{r}.
\end{equation}
\textbf{Recall that the overscript tilde notation denotes dimensionless or scaled quantities.} Unlike the gravitational point mass lens (Section \ref{sec:GravitationalPointMassLens}) and point diverging lens (Section \ref{sec:PointDivergingLens}), the Gaussian lens is a plasma lens, so we account for the effects of electromagnetic waves propagating through plasma via $f_{p}$ (Equation \ref{eq: plasma frequency}).

\begin{figure}
\plotone{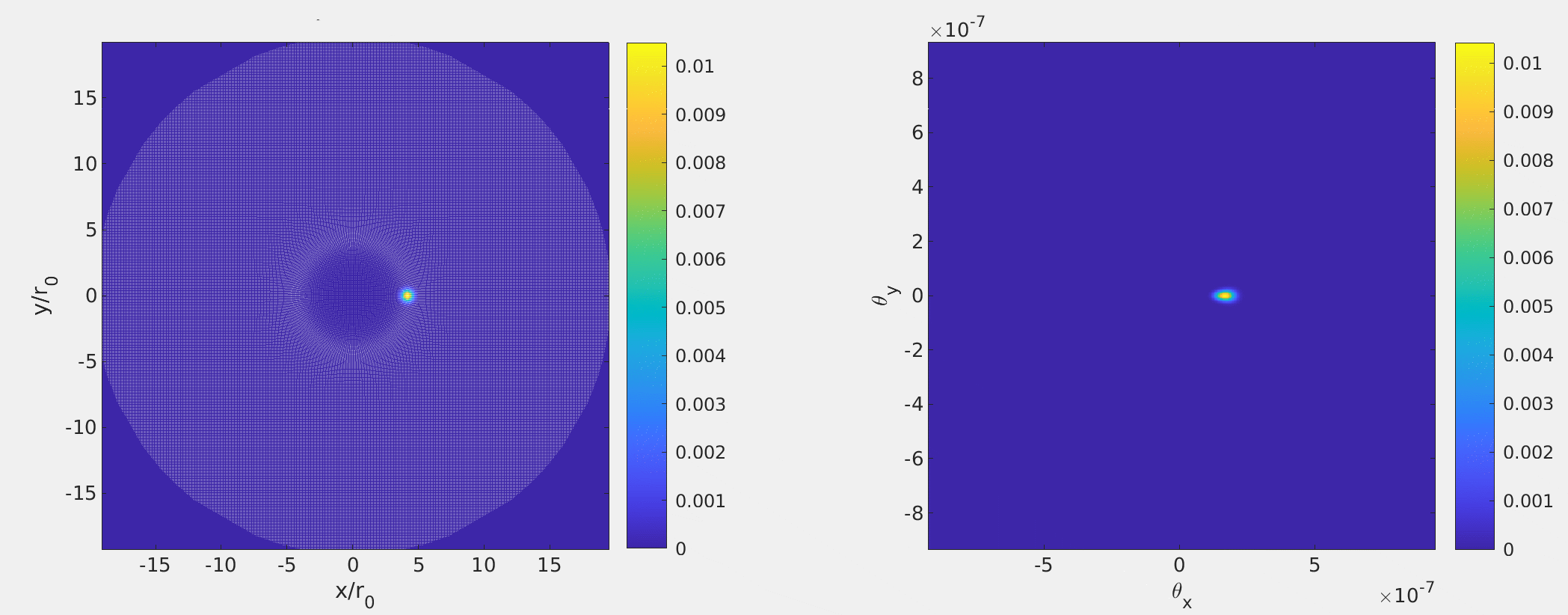}
\caption{Snapshot of a source with Gaussian-distributed brightness traversing horizontally across the source plane (left) and the resulting image on the observer plane (right) for a Gaussian plasma lens. Multiple-imaging results in an apparent stretching of the source on the image plane. The parameters are typical of a pulsar ESE with the lens placed halfway between observer and source: $D_{s} = 1 \, \textrm{kpc}$, $D_{d}=0.5 \, \textrm{kpc}$, $\rho_{H}=10^{-20}\,\textrm{g cm}^{-3}$, with a lens of length scale $r_{0}=\sigma=10 \, \textrm{AU}$. The colour bar indicates the normalized intensity.}
\label{fig:fluxMapsgaussiansigmalens1E1.png}
\end{figure}

\begin{figure}
\plotone{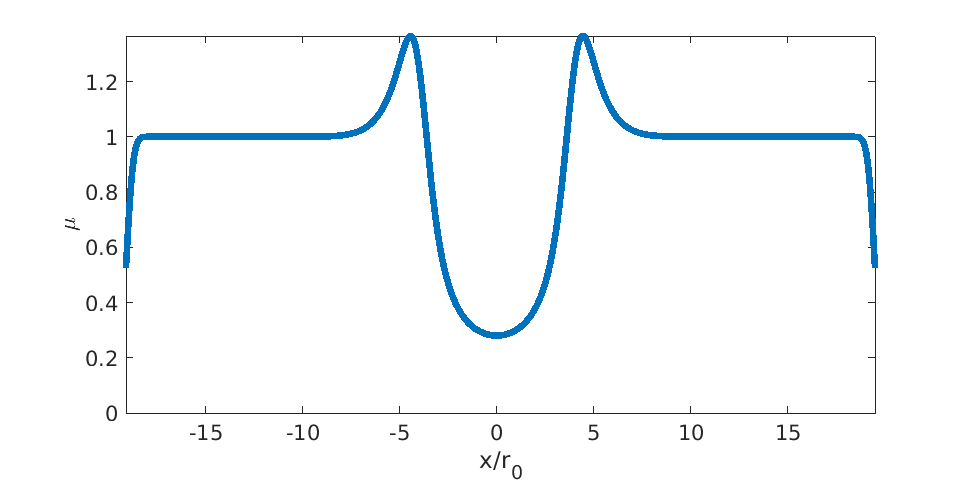}
\caption{Complete light curve as the source traverses horizontally from the left to right edge of the source plane (Figure \ref{fig:fluxMapsgaussiansigmalens1E1.png}, left panel).}
\label{fig:lightCurvegaussiansigmalens1E1_nosubtitle.png}
\end{figure}

\begin{figure}
\plotone{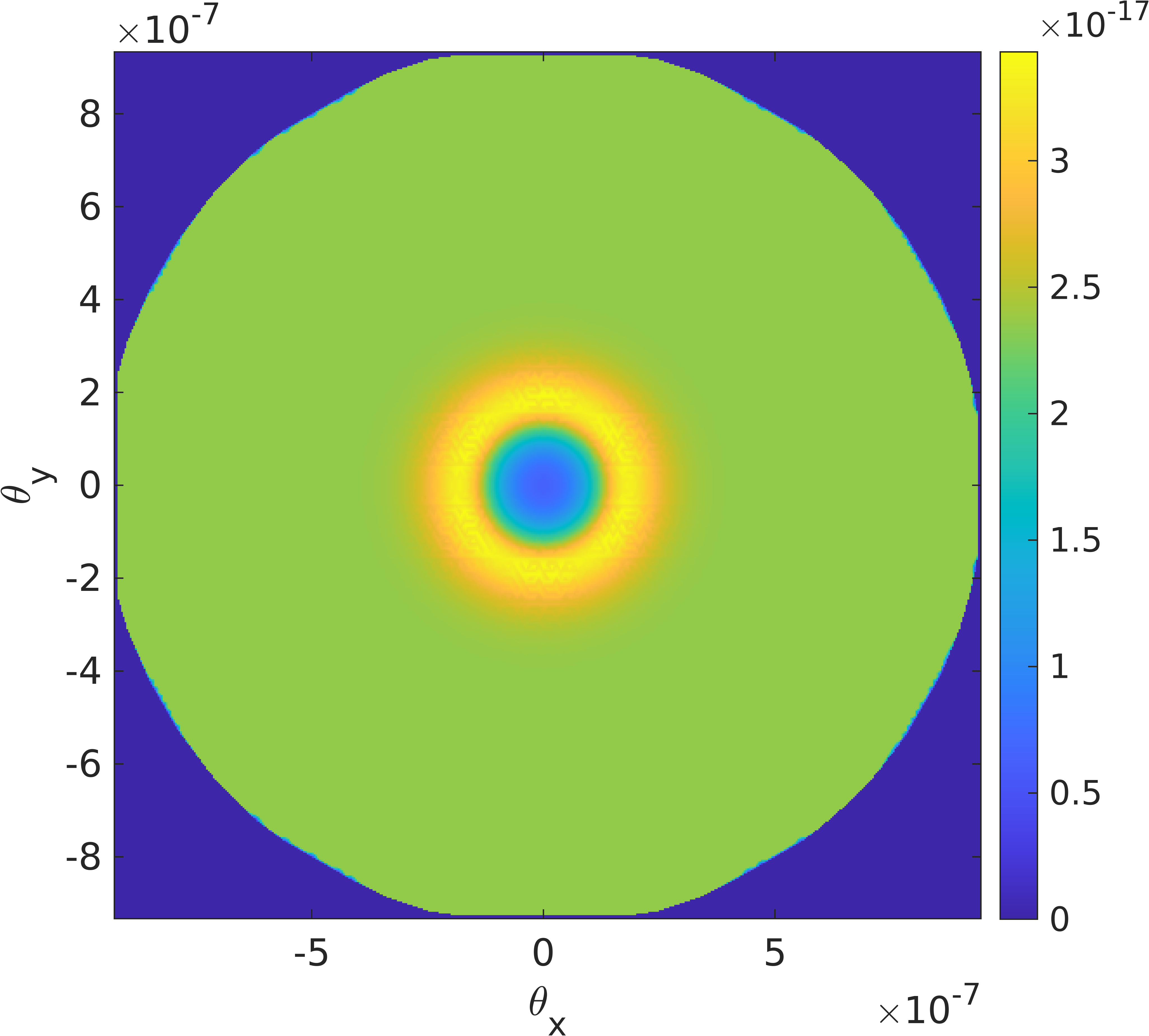}
\caption{Magnification map of the volumetric Gaussian lens example described in the text. Magnification is calculated by the ratio of Delaunay triangle areas on the image and source plane (Section \ref{sec:Magnification Maps}).}
\label{fig:magMapgaussiansigmalens1E1.png}
\end{figure}

Consider a system with a Gaussian lens placed halfway between an observer on Earth and a pulsar ESE event. The Gaussian lens has a length scale (standard deviation) $r_{0} = \sigma = 10 \, \textrm{AU}$, and hydrogen mass density $\rho_{H} = 10^{-20} \, \textrm{g cm}^{-3}$, typical of plasma lenses based on observations and models of ESEs \citep{Clegg1998,Stanimirovic2018}. Furthermore, pulsars in the Galaxy are located on the order $D_{s}=1\,\textrm{kpc}$ from the Earth, and the lens is placed halfway in between $D_{d}=0.5 \, \textrm{kpc}$. The resultant source and observer planes, and light curve are shown in Figures \ref{fig:fluxMapsgaussiansigmalens1E1.png} and \ref{fig:lightCurvegaussiansigmalens1E1_nosubtitle.png}, respectively, and a magnification map is shown in Figure \ref{fig:magMapgaussiansigmalens1E1.png}. A snapshot of the source is shown in the left panel of Figure \ref{fig:fluxMapsgaussiansigmalens1E1.png} with the corresponding observer image on the right panel of the same figure. White dots in the source plane (which are too small and numerous to be individually resolved in Figure \ref{fig:fluxMapsgaussiansigmalens1E1.png}, left panel) illustrate the correspondence of the pixels on the observer plane mapped to the source plane after ray-tracing and Delaunay triangulation (Section \ref{sec:Wavefront_Propagation_with_Delaunay_Triangulation}). Figure \ref{fig:lightCurvegaussiansigmalens1E1_nosubtitle.png} shows the complete normalized light curve after the source has travelled horizontally from the left edge of the source plane to the right edge.

\subsection{Rotating Magnetized Filament as a Plasma Lens}
\label{sec:Rotating Manetized Filament as a Plasma Lens}
\cite{Grafton2023MNRAS.522.1575G} constructed analytic models of rotating magnetized filaments observed in the extreme region of the Galaxy's Central Molecular Zone (CMZ) known as \textit{molecular tornadoes}. The formation of molecular tornadoes is theorized to arise from the shearing of a magnetic flux tube, which drives a torsional Alfv\'{e}n wave that propagates along the tube and triggers a cylindrical instability. Both axial (poloidal, $B_{z}$) and toroidal ($B_{\phi}$) magnetic fields thread the filament, which are characterized by the poloidal and toroidal flux-to-mass ratios formulated by \cite{Fiege1999a}
\begin{subequations}\label{eq:flux-to-mass ratios}
    \begin{equation}\label{eq:Gamma_z dimensional}
        \Gamma_{z} = \frac{B_{z}}{\rho},
    \end{equation}
    \begin{equation}\label{eq:Gamma_phi dimensional}
        \Gamma_{\phi}=\frac{B_{\phi}}{r\rho},
    \end{equation}
\end{subequations}
and taken as positive quantities. From the ideal magnetohydrodynamic (MHD) equations, \cite{Grafton2023MNRAS.522.1575G} found the radial mass density profile for the non-self-gravitating filament in equilibrium \citep[Equation 10]{Grafton2023MNRAS.522.1575G}
\begin{equation}\label{eq: Grafton2023 tilde{rho} density profile}
    \rho(r)=4 \pi \frac{-\sigma^{2} + [\sigma^{4} + \frac{P_{\infty}}{2\pi}(\Gamma_{z}^{2}+\Gamma_{\phi}^{2}r^{2})]^{1/2}}{\Gamma_{z}^{2}+\Gamma_{\phi}^{2}r^{2}},
\end{equation}
where $P_{\infty}$ is the total pressure including both gas and magnetic pressures everywhere within the filament and out to radial infinity, and $\sigma$ is the velocity dispersion. Here, we re-imagine the \cite{Grafton2023MNRAS.522.1575G} molecular tornado as a smaller scale rotating plasma filament.

Specifically, the case of constant $\Gamma_{\phi}$ and $\Gamma_{z}$ \citep[Section 3]{Grafton2023MNRAS.522.1575G} is the filamentary model used here as a plasma lens. The same length scale as \cite{Grafton2023MNRAS.522.1575G} is adopted where
\begin{equation}\label{eq: r0 length scale Grafton 2023}
    r_{0} = \frac{\Gamma_{z}}{\Gamma_{\phi}},
\end{equation}
with slight changes to the scaling relations between dimensionless quantities and physical quantities
\begin{equation}\label{eq:dimensionless scaling Gamma_z r rho}
\begin{split}
    \Gamma_{z} &= \frac{\sigma (4\pi)^{1/2}}{\rho_{0}^{1/2}} \tilde{\Gamma}_{z},  \\
    r &= r_{0} \tilde{r}, \\
    \rho &= \rho_{0} \tilde{\rho}.
\end{split}
\end{equation}
The scaling relations (Equation \ref{eq:dimensionless scaling Gamma_z r rho}) have free parameters $r_{0}$, $\rho_{0}$, and $\sigma$. ESE plasma lenses are on the order of $r_{0} \sim \textrm{AU}$, while typical molecular tornadoes have density $\rho_{0} = 10^{-21} \, \textrm{g cm}^{-3}$, and velocity dispersion $\sigma = 15 \, \textrm{km s}^{-1}$ \citep{Sofue2007,Rathborne2014,Au2017}. \cite{Au2017} previously explored a parameter space $10^{-3} \leq \tilde{\Gamma}_{z} \leq 100$ for their molecular tornado models. As a rough estimate, assuming $B \approx B_{z} = 0.4 \, \textrm{mG}$ \citep{Sofue2007}, $\rho_{0} = 10^{-21} \, \textrm{g cm}^{-3}$, and $\sigma=15 \, \textrm{km s}^{-1}$, Equations \ref{eq:Gamma_z dimensional} with \ref{eq:dimensionless scaling Gamma_z r rho} reveal $\tilde{\Gamma}_{z}$ is of order unity.

In dimensionless form, the radial mass density profile of the filament with constant $\Gamma_{z}$ and $\Gamma_{\phi}$ is \citep[Equation 21]{Grafton2023MNRAS.522.1575G}
\begin{equation}
    \tilde{\rho}(\tilde{r}) = \frac{-1 + [1+2\tilde{\Gamma}_{z}^{2}(1+\tilde{r}^{2})]^{1/2}}{\tilde{\Gamma}_{z}^{2}(1+\tilde{r}^{2})},
\end{equation}
which may be most conveniently scaled to a physical quantity via Equation \ref{eq:dimensionless scaling Gamma_z r rho} to calculate the refractive index (Section \ref{sec:Plasma Lens Examples}). The dimensionless gradient of the refractive index is
\begin{equation} \label{eq: dimensionless gradient n wills filament 21-Feb-2023}
    \vec{\tilde{\nabla}}n = \frac{1}{2\sqrt{1-(f_{p}/f)^{2}}} \left( \frac{f_{p}}{f} \right)^{2} \frac{1}{\tilde{\rho}} \frac{\partial \tilde{\rho}}{\partial \tilde{r}} \; \hat{r}
\end{equation}
where \citep[Equation 24]{Grafton2023MNRAS.522.1575G}
\begin{equation}
    \frac{\partial \tilde{\rho}}{\partial \tilde{r}} = - \frac{\tilde{\Gamma}_{z}^{2}\tilde{\rho}^{2}\tilde{r}}{1+\tilde{\Gamma}_{z}^{2}(1+\tilde{r}^{2})\tilde{\rho}}.
\end{equation}

\begin{figure}
\plotone{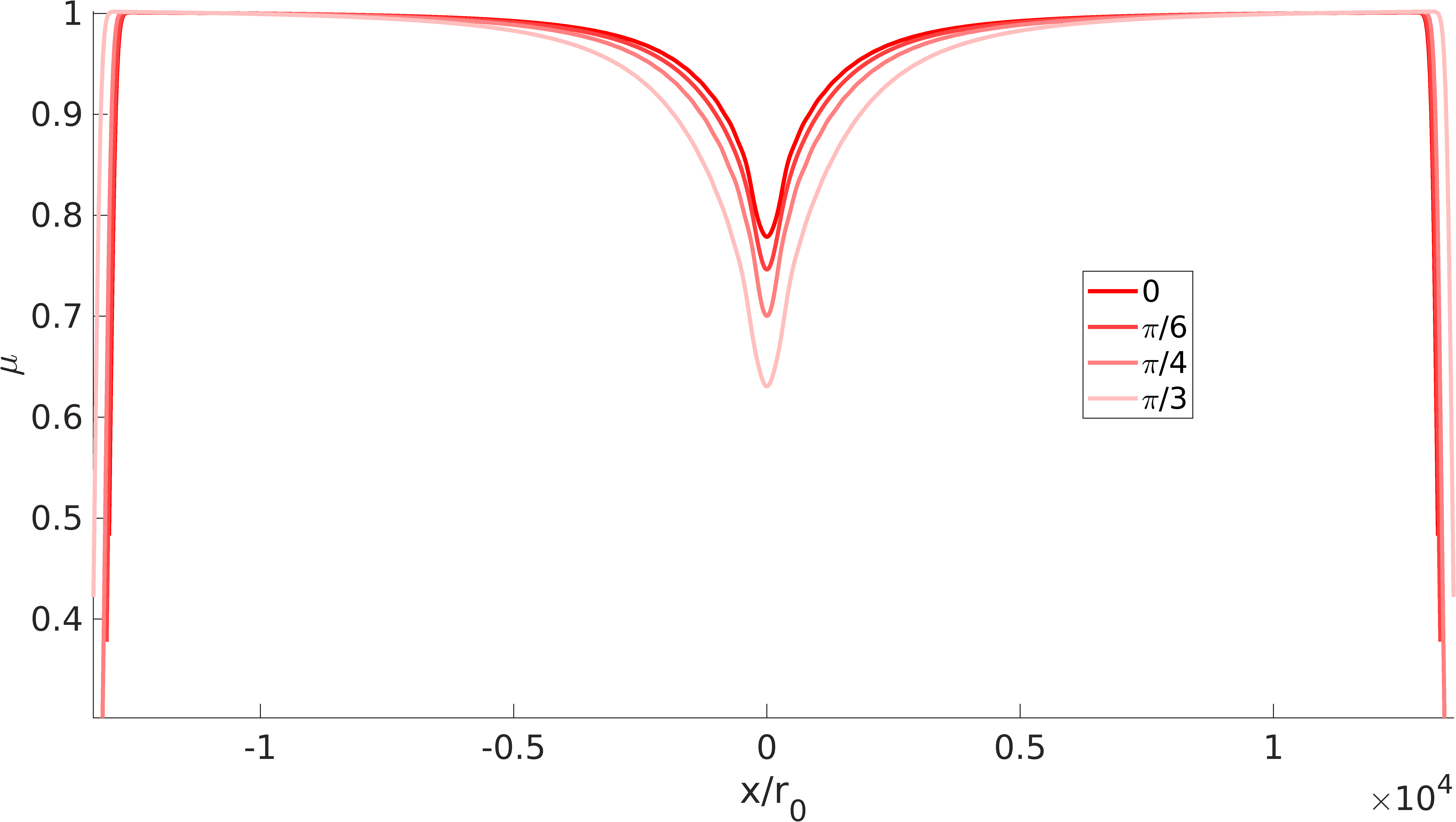}
\caption{Light curves for the \cite{Grafton2023MNRAS.522.1575G} filament reimagined as as plasma lens for inclination angles $i=[0, \pi/6, \pi/4, \pi/3]$ (shades of red). The source travels from left to right in the source plane, perpendicular to the long-axis of the filament when $i=0$, and takes the same path for each inclination angle. The lensing system is set up with the filament half way between the observer and source plane that $D_{s} = 1 \, \textrm{kpc}$, $D_{d} = 0.5 \, \textrm{kpc}$, with length scale $r_{0} = 10 \, \textrm{AU}$, core density $\rho_{0} = 10^{-21} \, \textrm{g cm}^{-3}$, and $\tilde{\Gamma}_{z} = 10^{-3}$.}
\label{fig:will_lightcurves.png}
\end{figure}

For example, consider a similar system to the Gaussian lens previously shown in Section \ref{sec:GaussianLens}, except now with the filament lens such that the length scale is $r_{0} = 10 \, \textrm{AU}$, $D_{s} = 1 \, \textrm{kpc}$, and $D_{d} = 0.5 \, \textrm{kpc}$. The filament is oriented such that its long axis runs vertically from top-to-bottom, perpendicular to the observer's perspective. In this orientation, the filament's inclination angle is defined as $i=0$. Rotating the top of the filament down such that the filament axis gradually becomes more aligned with the observer's line-of-sight corresponds to an increase in $i$. The resulting light curves are shown in Figure \ref{fig:will_lightcurves.png} for $\tilde{\Gamma}_{z} = 10^{-3}$.

\section{Discussion}
To explain the results shown in Section \ref{sec:Plasma Lens Examples}, consider Figures \ref{fig:fluxMapsgaussiansigmalens1E1.png} and \ref{fig:lightCurvegaussiansigmalens1E1_nosubtitle.png} for the Gaussian lens (Section \ref{sec:GaussianLens}). Some notable features include multiple-imaging seen in the observer plane in the right panel of Figure \ref{fig:fluxMapsgaussiansigmalens1E1.png}. The apparent radial stretching of the source on the observer plane due to multiple-imaging is similar to other divergent lens models \citep{Safonova2001PhRvD..65b3001S,Er2018,Rogers2019}. Notable in Figure \ref{fig:lightCurvegaussiansigmalens1E1_nosubtitle.png} are the caustic spikes around $x/r_{0}=5$ and dramatic drop in flux around $x/r_{0}=0$ which are signature features of ESEs. Comparing Figure \ref{fig:fluxMapsgaussiansigmalens1E1.png} with \ref{fig:lightCurvegaussiansigmalens1E1_nosubtitle.png}, it is easy to see that many pixels on the observer plane are mapped the same areas on the source plane as evidenced by the dense circular region of white point on the source plane centred on (0,0). When the source crosses these regions of dense points on the source plane, its brightness is mapped to multiple pixels on the observer plane, which results in multiple-imaging and an increase in brightness in the light curve. The opposite is true of the drop in the light curve in Figure \ref{fig:lightCurvegaussiansigmalens1E1_nosubtitle.png}; the source traverses through a region devoid of points on the source plane, which results in little to no source brightness mapped to the observer plane. This point is discussed further in Section \ref{sec:Practical Considerations}.

A lack of caustic spikes is notable in the light curves of Figure \ref{fig:will_lightcurves.png}. The light curves are mainly flat far away from the filament core and gradually decreasing more dramatically toward the filament core. Interestingly, these light curves resemble some of those produced by the filamentary plasma lens models in \cite{Rogers2020}. As the inclination angle increases and the filament axis becomes more aligned with the observer's line-of-sight, the drop in the light curve decreases more. Although these light curves do not possess the characteristic caustic spikes of ESE light curves, the relationship between an increasing inclination angle with a larger drop in the light curve is suggestive that the excessive density and pressure of plasma lenses are a geometrical effect \citep{Romani1987}. When the filament's long axis is more aligned with the observer, the greater deflection in the light rays away from the observer occurs as the rays travel through a region of greater electron column density.

It is also worth noting that the core densities used in the plasma lens models (Section \ref{sec:Plasma Lens Examples}) are still a few orders of magnitude in excess of typical ISM values. A wider exploration of the parameter space is underway to investigate possible models without excessive densities and produce ESEs (Section \ref{sec:Future Work}).

\subsection{Practical Considerations}
\label{sec:Practical Considerations}
As with many computational problems, computational speed is a limitation to the accuracy of the solution and practicality of solving the problem. Although physical hardware limitations do not necessarily affect the overall results and conclusions, it is important to recognize and to be cognizant of the limitations and consequences. For example, the memory limit is particularly important in choosing the number of rays and number of pixels. Generally, the calculation of identifying pixels on the observer plane with the Delaunay triangles they reside within (Section \ref{sec:Accounting_for_Multiple_Imaging}) is the most memory-intensive procedure. While this calculation is vectorized to increase the calculation speed, vectorization also increases memory use. There are a few consequences to limiting the number of rays and number of pixels. Qualitatively, the number of rays is most easily observed when the pixels on the observer plane are mapped to and shown on the source plane. When relatively fewer rays are traced, the points on the source plane are not as smoothly distributed. This is because fewer Delaunay triangles are created, resulting in larger Delaunay triangles containing more pixels per triangle, which show up as points distributed in triangular-shaped artifacts on the source plane, and some regions may be noticeably more densely populated with points than others. The resulting light curves may suddenly jump or drop in intensity.

Reducing the number of pixels reduces the number of points sampled by the source on the source plane, meaning less of its brightness mapped to the observer plane and accounted for in the calculation of the total intensity. In severe cases, there may be regions on the source plane which are devoid of points. When a significant portion of the source's total brightness (typically at or around the peak of the Gaussian brightness profile) is located in the gaps between points, the light curve shows a significant drop in intensity. If the source encounters several of these gaps then the light curve will appear ``bumpy'' or ``wavey''. The slight ringing effect seen in some curves of Figure \ref{fig:compare2analytic_divergentpointmass_N1e3_Ds0.1pc_Nx200} demonstrate this effect to a minor degree. This issue is largely mediated by increasing the number of pixels such that the source region is well populated with points, and to a lesser extent by the number of Delaunay triangles. For a fixed number of pixels residing within a Delaunay triangle on the observer plane, a larger triangle on the source plane can spread out the points more compared to a smaller triangle on the source plane. Additionally, increasing the size of the source's Gaussian brightness profile helps mediate this problem, and there are other consequences to consider concerning the source size, which are discussed next.

\subsection{Source Size}
\label{sec:Source Size}
Ideally, the source of the radio signal would be a point source to replicate a pulsar ESE scenario. However, as previously mentioned, practical limits to the number of pixels translates to the distribution and population of points on the source plane. If the source is smaller than the gaps between the points then the same problem arises, as previously discussed, where a ``bumpy'' light curve results.

In some situations, it is also difficult to tell whether some features in a light curve may be a natural consequence of the lensing model or artifacts due to a combination of the source size and finite number of points on the source plane (or pixels on the observer plane). For example, while generating light curves for the Gaussian plasma lens, some sets of parameters produced what seemed to be outer caustics like those shown in \cite{Clegg1998}. However, many of these light curves are riddled with artifacts due to the comparably small Gaussian brightness profile size of the source relative to the size of gaps between source points. It is possible that the overall shape of the light curve including the outer caustics are genuinely from the lens model, but the inability to confidently sample a small source size casts some doubt into the interpretation. The lack of outer caustics in our volumetric Gaussian lens light curve (Figure \ref{fig:lightCurvegaussiansigmalens1E1_nosubtitle.png}) could also be due to a genuine difference between \citeauthor{Clegg1998}'s (\citeyear{Clegg1998}) electron density model distributed as a two-dimensional Gaussian on the sky. Also of note is that features in the light curve such as caustics may be missed or smoothed out depending on the size of the source relative to the density of points on the source plane. \cite{Clegg1998} also remark and demonstrate for their model that the light curve \textit{does} change depending on the source size, however.

Despite the fact that the source can never be a perfect point, tests converge to known analytic solutions in the limit where the source becomes smaller and more point-like as previously shown in Section \ref{sec:TestCases}.

\subsection{Future Work}
\label{sec:Future Work}
The numerical method presented here lends itself well to ray-tracing through simulated volumetric turbulence, which has not yet been explored. Plasma lenses producing ESEs have so far been modeled as discrete structures. As mentioned previously (Section \ref{sec:Introduction}), it is theorized that ESEs due to TSIS may be the result of density fluctuations from turbulence. Often, the (possibly related) phenomenon of interstellar scattering and scintillation arcs are statistically modeled and/or by an effective scattering screen \citep[for example]{Cordes1985ApJ...288..221C,Cordes2016ApJ...817...16C,Pen2012MNRAS.421L.132P,Simard2018MNRAS.478..983S,Stinebring2001ApJ...549L..97S,Stinebring2022arXiv220708756S,Sprenger2022MNRAS.515.6198S,Baker2022arXiv221201417B,Jow2022MNRAS.514.4069J}. The eikonal ray-tracing method explored here opens the possibility for investigating three-dimensional volumetric turbulence as a lens. Efforts are currently underway to apply the numerical ray-tracing method presented here to density fluctuations following Kolmogorov turbulence, and simulated turbulence from PythonMHD, a Python-focused astrophysical MHD code under development at the University of Manitoba \citep{Leboe-McGowan2022}. Sheet-like structures are also being considered as their geometry may explain the overpressure problem when viewed edge-on (Section \ref{sec:Introduction}). A wider exploration of the parameter space for the Gaussian lens and filament lens inspired by \cite{Grafton2023MNRAS.522.1575G} is underway to understand the overpressure problem which has plagued plasma lens models so far.

\section{Conclusion}
Whereas prior studies have often utilised effective two-dimensional planes to study lensing, we have developed a numerical method to explore volumetric plasma lensing configurations that produce ESEs by solving the eikonal equation to trace ray trajectories. Rays start from a grid of pixel centres on the observer plane and are traced to the source plane to account for multiple imaging. Delaunay triangles connect adjacent rays to approximate the wavefront. The pixels are divided into subpixels, and assuming the barycentric coordinates are preserved throughout the trajectory, form a map between the observer and source planes. Gridded subpixels on the observer plane are mapped as points on the source plane. An intensity map is generated by mapping a Gaussian-distributed brightness profile on the source plane back to pixels on the observer plane. A light curve is generated by summing the total intensity contributed by all the pixels on the intensity maps generated as the source brightness profile moves across the source plane. Tests with known analytic solutions demonstrate convergence with our numerical solutions as the Gaussian source brightness distribution becomes more point-like. Examples of lensing models inspired by \cite{Clegg1998} and \cite{Grafton2023MNRAS.522.1575G} are shown with characteristic ESE light curve features in the former. Further investigations include exploring a larger parameter space for the examples presented here, sheet-like lenses, and volumetric turbulence.

\section*{Acknowledgements}
We acknowledge the support of the Natural Sciences and Engineering Research Council of Canada (NSERC), CGSD3-547708-2020. Cette recherche a \'{e}t\'{e} financ\'{e}e par le Conseil de recherches en sciences naturelles et en g\'{e}nie du Canada (CRSNG),  CGSD3-547708-2020. K.A. acknowledges the support of the University of Manitoba Graduate Fellowship (UMGF). J.D.F. acknowledges the support of a Discovery Grant from NSERC, RGPIN/5930-2017.


\bibliography{KJA1}{}
\bibliographystyle{aasjournal}



\end{document}